\documentclass[%
 reprint,
 superscriptaddress,
 amsmath,amssymb,
 aps,
longbibliography,
prl,
floatfix,	
]{revtex4-1}

\usepackage{amssymb,amsmath, setspace}
\usepackage{dcolumn}% Align table columns on decimal point
\usepackage{bm}% bold math
\usepackage{threeparttable}

\usepackage{ifpdf}
\usepackage{cancel}
\ifpdf
 \usepackage[pdftex]{graphicx}
 \DeclareGraphicsExtensions{.pdf}
  \usepackage[hyperindex=true]{hyperref}
\else
  \usepackage{graphicx}
  \DeclareGraphicsExtensions{.ps,.eps}

\usepackage[a4paper,dvipdfm,hyperindex=true]{hyperref}
\fi

\usepackage{color}

\def\NNO{NdNiO$_2$}

\def\LNO{LaNiO$_2$}
\def\CCO{CaCuO$_2$}
\def\dxtyt{$d_{x^2-y^2}$}
\def\dzt{$d_{z^2}$}

\begin{document}
\preprint{DRAFT 1}
\title[{\em Ab initio} wavefunction analysis ... ]{{\em Ab initio} wavefunction analysis of electron removal quasi-particle state of NdNiO$_2$ with fully correlated quantum chemical methods
} 

\author{Vamshi M. Katukuri}

 \email{V.Katukuri@fkf.mpg.de}%
 \affiliation{%
 	Max Planck Institute for Solid State Research, Heisenbergstrasse 1, 70569 Stuttgart, Germany
 }%
\author{ Nikolay A. Bogdanov}%
\affiliation{%
	Max Planck Institute for Solid State Research, Heisenbergstrasse 1, 70569 Stuttgart, Germany
}%
% \email{N.Bogdanov@fkf.mpg.de}
%\author{Liviu Hozoi}
%\affiliation{IFW Dresden, Helmholtzstrasse 20, 01069 Dresden, Germany}

%\author{Jeroen van den Brink}
%\affiliation{IFW Dresden, Helmholtzstrasse 20, 01069 Dresden, Germany}
%\affiliation{Department of Physics, Technical University Dresden, Helmholtzstraße 10, 01069 Dresden, Germany}

\author{Ali Alavi}%
\email{A.Alavi@fkf.mpg.de}%
\affiliation{%
 Max Planck Institute for Solid State Research, Heisenbergstrasse 1, 70569 Stuttgart, Germany
}%
\affiliation{%
Department of Chemistry, University of Cambridge, Lensfield Road, Cambridge CB2 1EW, UK
}

\date{\today}

\begin{abstract}

The discovery of superconductivity in hole-doped infinite-layer NdNiO$_2$ --- a transition metal (TM)
oxide that is both isostructural and isoelectronic to cuprate superconductors --- 
 has lead to renewed enthusiasm in the hope of understanding the origin of unconventional
superconductivity. Here, we investigate the electron-removal states in infinite-layered Ni$^{1+}$
oxide, NdNiO$_2$, which mimics hole-doping, with the state-of-the-art many-body multireference quantum chemistry methods. 
From the analysis of the many-body wavefunction we find that the hole-doped $d^8$ ground state of NdNiO$_2$ is very different from the $d^8$ ground state in isostructural cuprate analog CaCuO$_2$, although the parent $d^9 $ ground states are for the most part identical.
We show that the doped hole in NdNiO$_2$ mainly localizes on the Ni 3\dxtyt\ orbital to form a closed-shell singlet, and this singlet configuration contributes to $\sim$40\% of the wavefunction. 
In contrast, in CaCuO$_2$ the Zhang-Rice singlet configurations contribute to $\sim$65\% of the wavefunction. 
With the help of the quantum information concept of entanglement entropy, we quantify the different types
of electronic correlations in the nickelate and cuprate compounds, and find that the dynamic radial-type correlations within the Ni $d$ manifold are persistent in hole-doped NdNiO$_2$. 
As a result, the $d^8$ multiplet effects are stronger and the additional hole foot-print is more three dimensional in NdNiO$_2$. 
Our analysis shows that the most commonly used three-band Hubbard model employed to express the doped scenario in cuprates represents $\sim$90\% of the $d^8$ wavefunction for \CCO, but such a model grossly approximates the $d^8$ wavefunction for NdNiO$_2$ as it only stands for $\sim$60\% of the wavefunction. 

\end{abstract}
\maketitle

\section{Introduction}
For more than three decades, understanding the mechanism of superconductivity observed at high critical temperature (HTC) in
strongly correlated cuprates~\cite{LaCuO2_Bednorz_86} has been the ``holy grail” 
of many theoretical and experimental condensend matter researchers.
In this context, the observation of superconductivity in 
 nickelates $Ln$NiO$_2$, $Ln$=\{La, Nd and Pr\} ~\cite{li_superconductivity_2019,osada_superconducting_2020,osada_nickelate_2021} upon doping with holes is remarkable. 
 These superconducting nickelates are isostructural as well as isoelectronic to 
HTC cuprate superconductors and thus enable the comparison of
the essential physical features that may be playing a crucial role in the mechanism driving superconductivity.

$Ln$NiO$_2$ family of compounds are synthesized in the so-called infinite-layer structure, where NiO$_2$ and $Ln$ layers are stacked alternatively~\cite{li_superconductivity_2019}. 
The NiO$_2$ planes are identical to the CuO$_2$ planes in HTC cuprates which host much of the physics leading to superconductivity~\cite{keimer_quantum_2015}. 
A simple valence counting of the these nickelates reveals a {1+} oxidation state for Ni ({2-} for O and {3+} for $Ln$) with 9 electrons in the $3d$ manifold. 
In the cuprates, the Cu$^{2+}$ oxidation state gives rise to the same $3d^9$ electronic configuration.
Contrary to many nickel oxides where the Ni atom sits in an octahedral cage of oxygens, in the infinite-layered structure, square planar NiO$_4$ plaques are formed without the apical oxygens. 
The crystal field due to square-planar oxygen coordination stabilizes the $d_{z^2}$ orbital of the $e_g$ manifold, making its energy close to the $t_{2g}$ orbitals (the $3d$ orbitals split to 3-fold $t_{2g}$ and 2-fold $e_g$ sub-shells in an octahedral environment). With $d^9$ occupation, a half-filled $d_{x^2-y^2}$-orbital system is realized as in cuprates.
In fact, recent resonant inelastic X-ray scattering (RIXS) experiments~\cite{rossi2020orbital} as well as the {\it ab initio} correlated multiplet calculations~\cite{katukuri_electronic_2020} confirm that the Ni$^{1+}$ $d$-$d$ excitations in \NNO\ are similar to the Cu$ ^{2+} $ ions in cuprates~\cite{moretti_sala_energy_2011}.

 Several electronic structure calculations based on density-functional theory (DFT) have shown that in monovalent nickelates the Ni 3$d_{x^2-y^2}$ states sit at the Fermi energy level ~\cite{lee_infinite-layer_2004,liu_electronic_njpqm_2020,zhang_effective_prl_2020}.
 These calculations further show that the nickelates are more close to the Mott-Hubbard insulating limit with a decreased Ni  $3d$- O $2p$ hybridization compared to cuprates. 
 The latter are considered to be charge transfer insulators~\cite{zsa_mott_charge_transfer_1985} where excitations across the electronic band gap involves O $2p$ to Cu $3d$ electron transfer.
Correlated wavefunction-based calculations~\cite{katukuri_electronic_2020} indeed find that the contribution from the O $2p$ hole configuration to the ground state wavefunction in \NNO\ is four times smaller than in the cuprate analogue \CCO.
 X-ray absorption and photoemission spectroscopy experiments~\cite{hepting2020a,goodge-a} confirm the Mott behavior of nickelates. 
 
In the cuprate charge-transfer insulators, the strong hybridization of the Cu 3\dxtyt\ and O $2p$ orbitals result in O $2p$ dominated bonding and Cu 3\dxtyt\ -like antibonding orbitals. As a consequence, the doped holes primarily reside on the bonding O $2p$ orbitals, making them singly occupied. 
The unpaired electrons on the Cu \dxtyt\ and the O $2p$ are coupled antiferromagnetically resulting in the famous Zhang-Rice (ZR) spin singlet state~\cite{zhang_effective_1988}. 
In the monovalent nickelates, it is unclear where the doped-holes reside. Do they form a ZR singlet as in cuprates? Instead, if the holes reside on the Ni site, do they form a high-spin local triplet  with two singly occupied Ni $3d$ orbitals and aligned ferromagnetically or a low-spin singlet with either both the holes residing in the Ni 3\dxtyt\ orbital or two singly occupied Ni 3$d$ but aligned anti-parallel.
% As a result of decreased $3d$-$2p$ hybridization or an increased charge transfer gap ($\delta_{CT}$), the character of doped holes is . In the cuprates classified as a charge-transfer insulator in the Zaanen-SawatzkyAllen phase diagram [44], the doped holes mainly enter the oxygen 2p orbitals because the charge-transfer energy ∆dp is smaller than the 3d-orbital Hubbard interaction Udd. 
While Ni L-edge XAS and RIXS measurements~\cite{rossi2020orbital} conclude that an orbitally polarized singlet state is predominant, where doped holes reside on the Ni 3\dxtyt\ orbital, O K-edge electron energy loss spectroscopy~\cite{goodge-a} reveal that some of the holes also reside on the O $2p$ orbitals. 
On the other hand, calculations based on multi-band $d-p$ Hubbard models show that the fate of the doped holes is determined by a subtle interplay of Ni onsite ($U_{dd}$), Ni $d$ - O $2p$ inter-site ($U_{dp}$) Coulomb interactions and the Hund's coupling along with the charge transfer gap~\cite{jiang_critical_prl_2020,Plienbumrung_condmat_2021}. 
However, with the lack of extensive experimental data, it is difficult to identify the appropriate interaction parameters for a model Hamiltonian study, let alone identifying the model that best describes the physics of superconducting nickelates.
 %However, this feature is absent in \LNO, where superconductivity is discovered more recently~\cite{osada_nickelate_2021}. 
 
Despite the efforts to discern the similarities and differences between the monovalent nickelates and superconducting cuprates, there is no clear understanding on the nature of doped holes in NdNiO$_2$.
Particularly, there is no reliable parameter-free \textit{ab initio} analysis of the hole-doped situation. 
In this work, we investigate the hole-doped ground state in \NNO\ and draw parallels to the hole doped ground state of cuprate analogue \CCO. 
We use fully {\it ab initio} many-body wavefunction-based quantum chemistry methodology % based on the construction of configuration interaction (CI) wavefunction
to compute the ground state wavefunctions for the hole doped \NNO\ and \CCO. 
We find that the doped hole in NdNiO$_2$ mainly localizes on the Ni 3\dxtyt\ orbital to form a closed-shell singlet, and this singlet configuration contributes to $\sim$40\% of the wavefunction. 
In contrast, in CaCuO$_2$ the Zhang-Rice singlet configurations contribute to $\sim$65\% of the wavefunction. 
The persistent dynamic radial-type correlations within the Ni $d$ manifold result in stronger $d^8$ multiplet effects than in \CCO,
and consequently the additional hole foot-print is more three-dimensional in NdNiO$_2$. 
Our analysis shows that the most commonly used three-band Hubbard model to express the doped scenario in cuprates represents ~ 90\% of the $d^8$ wavefunction for \CCO, but such a model grossly approximates the $d^8$ wavefunction for the NdNiO$_2$ as it only stands for $\sim$60\% of the wavefunction.
%While much of the research work was focused on NdNiO$_2$ where superconductivity was first observed, it is not clear how the electronic structures of \LNO\ and \PNO\ compare with \NNO. Although the main crystal motif -- NiO$_4$ planes-- remain the same, the difference in the size of the rare-earth ions results in small but important changes in the Ni-O distances, see Table 1. 

%In this work we compute the $d$-$d$ excitations 

% For Original Research articles, please note that the Material and Methods section can be placed in any of the following ways: before Results, before Discussion or after Discussion.

In what follows, we first describe the computational methodology we employ in this work where we highlight the novel features of the methods and provide all the computational details.
We then present the results of our calculations and conclude with a discussion. 

\section{The wavefunction quantum chemistry method}
{\it Ab initio} configuration interaction (CI) wavefunction-based quantum chemistry methods, particularly
the post Hartree-Fock (HF) complete active space self-consistent field (CASSCF) and the multireference perturbation theory (MRPT), are employed.
These methods not only facilitate systematic inclusion of electron correlations, but also enable to quantify different types of correlations, static vs dynamic~\cite{helgaker_molecular_2000}.  
These calculations do not use any \textit{ad hoc} parameters to incorporate electron-electron interactions unlike other many-body methods, instead, they are computed fully {\it ab initio} from the kinetic and Coulomb integrals. 
Such \textit{ab initio} calculations provide techniques to systematically analyze electron correlation effects and offer insights into the electronic structure of correlated solids that go substantially beyond standard DFT approaches, e.g., see Ref.~\cite{Munoz_afm_htc_qc_prl_2000,CuO2_dd_hozoi11,book_Liviu_Fulde,Bogdanov_Ti_12,katukuri_electronic_2020} for the $ 3d $ TM oxides and Ref.~\cite{katukuri_PRB_2012,Os227_bogdanov_12,213_rixs_gretarsson_2012,Katukuri_ba214_prx_2014,Katukuri_njp_2014} for $ 5d $ compounds.
%The larger active spaces considered in the present work are at the limit of what can be achieved today, and allow us to not only capture all the static and large portion of dynamic correlations but also enable us to understand their significance.  
\subsection{Embedded cluster approach}
Since strong electronic correlations are short-ranged in nature \cite{fulde_new_book}, a local approach for the calculation of the $N$ and $N\pm$1 –electron wavefunction is a very attractive option for transition metal compounds. 
In the embedded cluster approach, a finite set of atoms, we call quantum cluster (QC), is cut out from the infinite solid and many-body quantum chemistry methods are used to calculate the electronic structure of the atoms within the QC. 
The cluster is ``embedded” in a potential that accounts for the part of the crystal that is not treated explicitly.
In this work,  we represent the embedding potential with an array of point charges (PCs) at the lattice positions that are fitted to reproduce the Madelung crystal field in the cluster region~\cite{ewald}.
Such procedure enables the use of quantum chemistry calculations for solids  involving transition-metal or lanthanide ions, see Refs.~\cite{katukuri_ab_2012,katukuri_electronic_2014,babkevich_magnetic_2016}.

%For computing the $d$-$d$ excitations and the ground state wavefunction analysis, we used a quantum cluster (Q1) consisting of five NiO$ _{4} $ square plaques along with eight neighboring La (Pr) ions. 
%To obtain the superexchange coupling, two-magnetic-site quantum clusters (Q2) were considered. 
%The Q2 clusters included two NiO$_4$ square units, six neighboring Ni$ ^{1+} $  ions and all adjacent La (Pr)$ ^{3+} $ ions. 
%The Ni atoms neighboring the clusters Q1 and Q2 are also included in the cluster.  

%For $J$ calculations, we used all electron cc-pVTZ and cc-pVQZ basis sets for central Ni and bridging O ions~\cite{dunning_jr_gaussian_1989,balabanov_systematically_2005}, while the other oxygens are represented with cc-pVDZ basis sets~\cite{balabanov_systematically_2005}.
%Large-core effective potentials were employed for other species as for $d$-$d$ excitation calculations.

\subsection{Complete active space self-consistent field}
% Starting from a set of orbitals obtained from a single-reference mean-field calculation, the orbitals are further optimized with respect to a multireference wavefunction. 
% This results in a self consistent set of single particle orbital basis in which the CI wavefunction can be expanded. 
% To this end, we use the CASSCF approach. 
CASSCF method~\cite{book_QC_00} is a specific type of multi-configurational (MC) self-consistent field technique in which a complete set of Slater determinants or configuration state functions (CSFs) is used in the expansion of the CI wavefunction is defined in a constrained orbital space, called the active space. 
%This is achieved by dividing the orbitals in a given system into three subspaces: 
%(a) inactive space where all orbitals are doubly occupied, 
%(b) active space, within which a full-CI expansion is considered, and 
%(c) virtual space, consists of orbitals that are kept unoccupied.
In the CASSCF(n,m) approach, a subset of $n$ active electrons are
fully correlated among an active set of $m$ orbitals, leading to a highly multi-configurational (CAS) reference wavefunction.
%The electrons and orbitals included in the active space are those that contribute most to the multireference character of the particular system of study,  
%and their occupation will be a non-integer number between 0 and 2.   
CASSCF method with a properly chosen active space guarantees a qualitatively correct wavefunction for strongly correlated systems where static correlation~\cite{book_QC_00} effects are taken into account. 
%A single reference wavefunction based methods fail completely in this situation. 
 %
We consider active spaces as large as CAS(24,30) in this work. 
%For the $d^9$ ground state wavefunction analysis and $d$-level excitation calculations, a large active space  
%consisting of five Ni  3$d$, all the O 2$p$ orbitals of the NiO plaque and the corresponding so-called ``double-shell" orbitals (Ni 4$d$ and O 3$p$) plus the semi-core Ni (Cu) 3$s$ and unoccupied 4$s$ orbitals, was considered resulting in a  35 electrons in 36 orbitals -- CAS(35,36)SCF -- correlated calculation. 
%The NN Ni$^{1+}$ ions were approximated with closed shell Cu$^{1+}$ ions to avoid spin-couplings with the central Ni$^{1+}$ (Cu$^{2+}$) ion, and the doubly occupied Cu 3\dxtyt\ orbitals were frozen at the HF level. 
%For computing the $d^8$ wavefunction and its multiplet structure, all five Ni$^{1+}$ are kept open-shell. 
%The active space in this case consisted of additional 20 orbital and 15 electrons comprising of five 3$d$ orbitals and 9 electrons from each of the NN Ni$^{1+}$ ions, making it a CAS(50,56)SCF calculation.
%
%The exchange coupling in the $d^9$ oxides is primarily of the superexchange type that depends on the virtual hopping of electrons (or holes) through the bridging oxygen and the effective on-site Coulomb repulsion ($U_{\rm eff}$) on the $3d$ orbitals of the Ni ions.
%To describe these two process accurately, we incorporated all the Ni 3$d$ and the corresponding double-shell $4d$ (10 + 10) orbitals as well as the bridging oxygen $2p$ and $3p$ (3 + 3) orbitals in the CASSCF active space. 
%
Because the conventional CASSCF implementations based on deterministic CI space (the Hilbert space of all possible configurations within in the active space) solvers are limited to active spaces of 18 active electrons in 18 orbitals,
%In the last decade, there has been a steady development of several new and efficient algorithms for solving the CI problem (CI solvers).  
we use the full configuration interaction quantum Monte Carlo (FCIQMC)~\cite{booth_fermion_2009,cleland_survival_2010,guther_neci_2020} and density matrix renormalization group (DMRG) theory~\cite{chan_density_2011,sharma_spin-adapted_2012} algorithms to solve the eigenvalue problem defined within the active space. 
%We used both DMRG~\cite{sharma_spin-adapted_2012} and FCIQMC~\cite{booth_fermion_2009,guther_neci_2020} methods to solve the eigenvalue problem defined within the active space. 
%as conventional deterministic solvers are incapable to handle the resulting large Hilbert spaces. 
%The number of renormalized states ($M$) was set to 3000 to guarantee convergence of the total energies. 
%We employed {\sc pyscf} quantum chemistry package~\cite{sun_pyscf_2017} for all the calculations.   

\subsection{Multireference perturbation theory}
While the CASSCF calculation provides a qualitatively correct wavefunction, for a quantitative description of a strongly correlated system, dynamic correlations~\cite{book_QC_00} (contributions to the wavefunction from those configurations related to excitations from inactive to active and virtual, and active to virtual orbitals) are also important and must be accounted for.  
A natural choice is variational multireference CI (MRCI) approach where the CI wavefunction is extended with excitations involving orbitals that are doubly occupied and empty in the reference CASSCF wavefunction \cite{book_QC_00}. 
An alternative and computationally less demanding approach to take into account dynamic correlations is based on perturbation theory in second- and higher-orders.  
In multireference perturbation theory (MRPT) MC zeroth-order wavefunction is employed and excitations to the virtual space are accounted by means of perturbation theory.  
If the initial choice of the MC wavefunction is good enough to capture the large part of the correlation energy, then the perturbation corrections are typically small. 
The most common variations of MRPT are the complete active space second-order perturbation theory (CASPT2)~\cite{anderson_caspt2_1992} and the n-electron valence second-order perturbation theory (NEVPT2)~\cite{angeli_nevpt2_2001} which differ in the type of zeroth-order Hamiltonian $H_0$ employed.
%While the CASPT2 uses effective one-electron Fock-type operator as $H_0$~\cite{anderson_caspt2_1992}, $H_0$ in the NEVPT2 includes
%two-electron operator restricted to the active space, as proposed by Dyall~\cite{dyall_1995}.
%A more recent variant multireference linearized coupled-cluster theory (MRLCC)~\cite{sharma_multireference_2015} uses Fink's Hamiltonian~\cite{FINK2006461} 
%and is highly efficient when large active space reference wavefunctions are used. 
%The latter two methods outperform CASPT2 method as they do not suffer from the intruder state problems associated with CASPT2~\cite{anderson_caspt2_1992}.

\begin{figure}[!t]
	\begin{center}
		\includegraphics[width=0.450\textwidth]{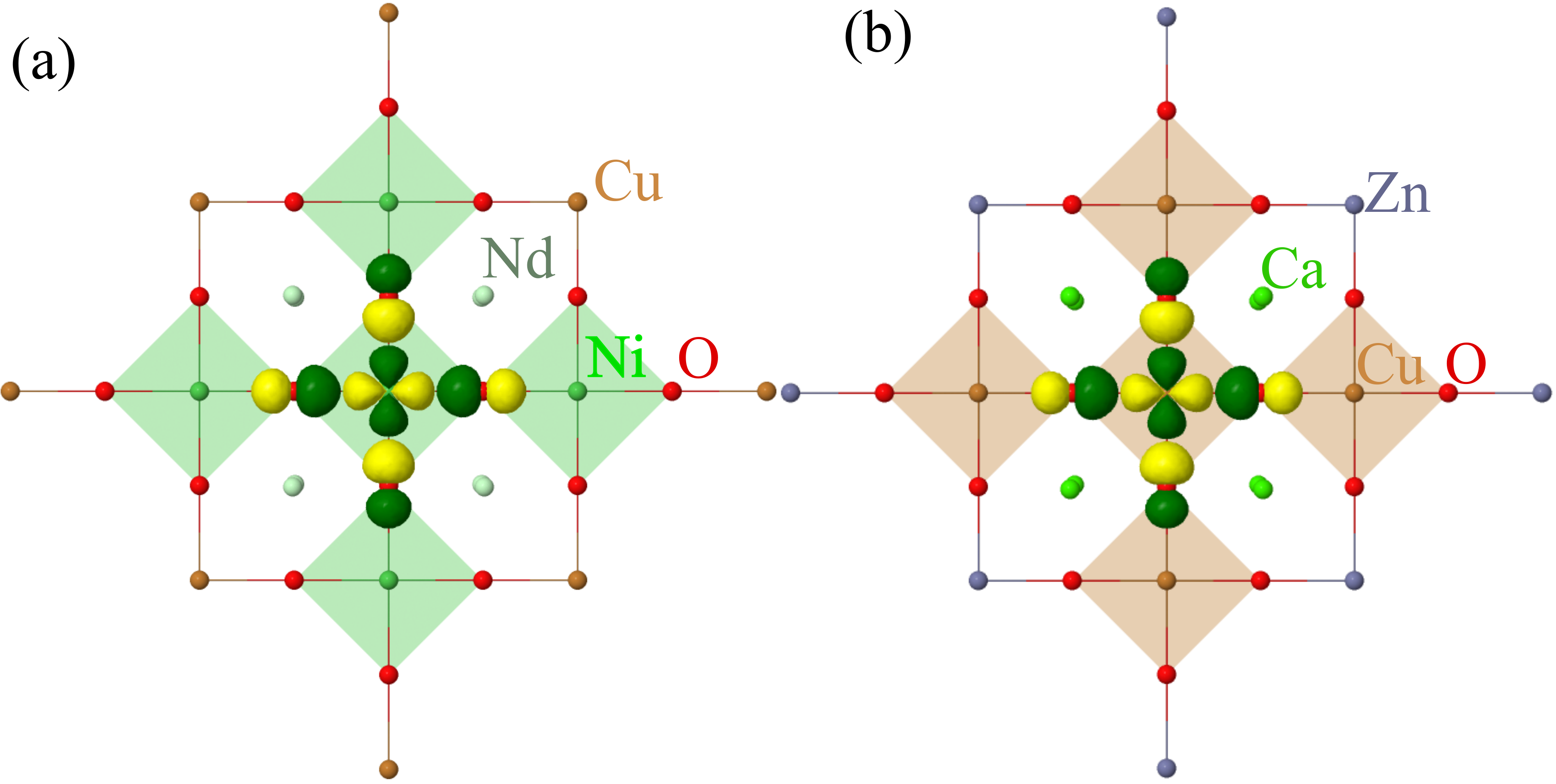}	
		\caption{Quantum cluster of five NiO$_4$ (a) and CuO$_4$ (b) plaques considered in our calculations.  The point-charge embedding is not shown. 
			The symmetry adapted localized 3\dxtyt\ and the  oxygen Zhang-Rice-like  2$p$ orbitals, the  basis in which the wavefunction in Table~\ref{wfn} is presented are shown in yellow and green color. }
		\label{fig1}
	\end{center}
\end{figure}

\section{The {\em ab initio} model}
Before we describe the {\em ab initio} model we consider, let us summarize the widely used and prominent model Hamiltonian to study the nature of doped hole in HTC cuprates and also employed for monovalent nickelates lately. 
It is the three-band Hubbard model~\cite{emery_3b_hubbard_prl_1987} with 
three orbital degrees of freedom (bands) which include the $d$ orbital of Cu with $x^2-y^2$ symmetry and the in-plane oxygen $p$ orbitals aligned in the direction of the nearest Cu neighbours. 
These belong to the $b_1$ irreducible representation (irrep) of the $D_{4h}$ point group symmetry realized at the Cu site of the CuO$_4$ plaque, the other Cu $d$ orbitals belong to $a_1$  ($d_{z^2}$), $b_2$ ($d_{xy}$) and $e$  ($d_{xz,yz}$) irreps.
The parameters in this Hamiltonian include the most relevant hopping and Coulomb interactions within this set of orbitals.  
More recently, the role of the Cu $3d$ multiplet structure on the hole doped ground state is also studied~\cite{jiang_cuprates_prb_2020}.  
While this model explains certain experimental observations, there is still a huge debate on what is the minimum model to describe the low-energy physics of doped cuprates. 
Nevertheless, this model has also been employed to investigate the character of the doped hole in monovalent nickelates~\cite{jiang_critical_prl_2020,Plienbumrung_condmat_2021, Plienbumrung_prb_2021}.  

Within the embedded cluster approach described earlier, 
we consider a QC of five NiO$_4$ (CuO$_4$) plaques that includes five Ni (Cu) atoms, 16 oxygens and 8 Nd (Ca) atoms. The 10 Ni (Cu) ions neighbouring to the cluster are also included in the QC, however, these are considered as total ion potentials (TIPs). % without any basis functions. 
The QC is embedded in point charges that reproduce the electrostatic field of the solid environment.
We used the crystal structure parameters for the thin film samples reported in Ref.~\cite{li_superconductivity_2019,hayward_synthesis_2003,kobayashi_compounds_1997,karpinski_single_1994}.

We used all-electron atomic natural orbital (ANO)-L basis sets of tripple-$\zeta$ quality with additional polarization functions -- [$7s6p4d2f1g$] for Ni (Cu)~\cite{roos_new_2005} 
and [$4s3p2d1f$] for oxygens~\cite{roos_main_2004}.
For the eight Nd (Ca) atoms large core effective potentials~\cite{dolg_energy-adjusted_1989,dolg_combination_1993,kaupp_pseudopotential_1991} and associated [$3s2p2d$] basis functions were used. 
In the case of Nd, the $f$-electrons were incorporated in the core.   
Cu$ ^{1+} $ (Zn$^{2+}$) total ion potentials (TIPs) with [$2s1p$] functions were used for the 10 Ni$^{1+}$ (Cu$^{2+}$)~\cite{ingelmann_thesis}~\footnote{Energy-consistent Pseudopotentials of Stuttgart/Cologne group, \url{http://www.tc.uni-koeln.de/cgi-bin/pp.pl?language=en,format=molpro,element=Zn,job=getecp,ecp=ECP28SDF}, [Accessed: 15-Sept-2021]}
neighbouring ions of the QC. 
%to avoid the spin-couplings of the Ni$^{1+}$ ions in the Q1 cluster with the neighboring Ni$ ^{1+} $ ions we replaced them with closed-shell Cu$ ^{1+} $ ions total ion potentials.
%We found that basis functions on top of the TIPs were essential for representing the charge redistribution on these ions to reproduce correctly the crystal field splittings.   
%Test calculations were performed to estimate the scalar relativistic effects within Douglas-Kroll-Hess integral correction and no significant deviations were found and hence they were not included in the data reported. 

\begin{figure}[!t]
	\begin{center}
		\includegraphics[width=0.480\textwidth]{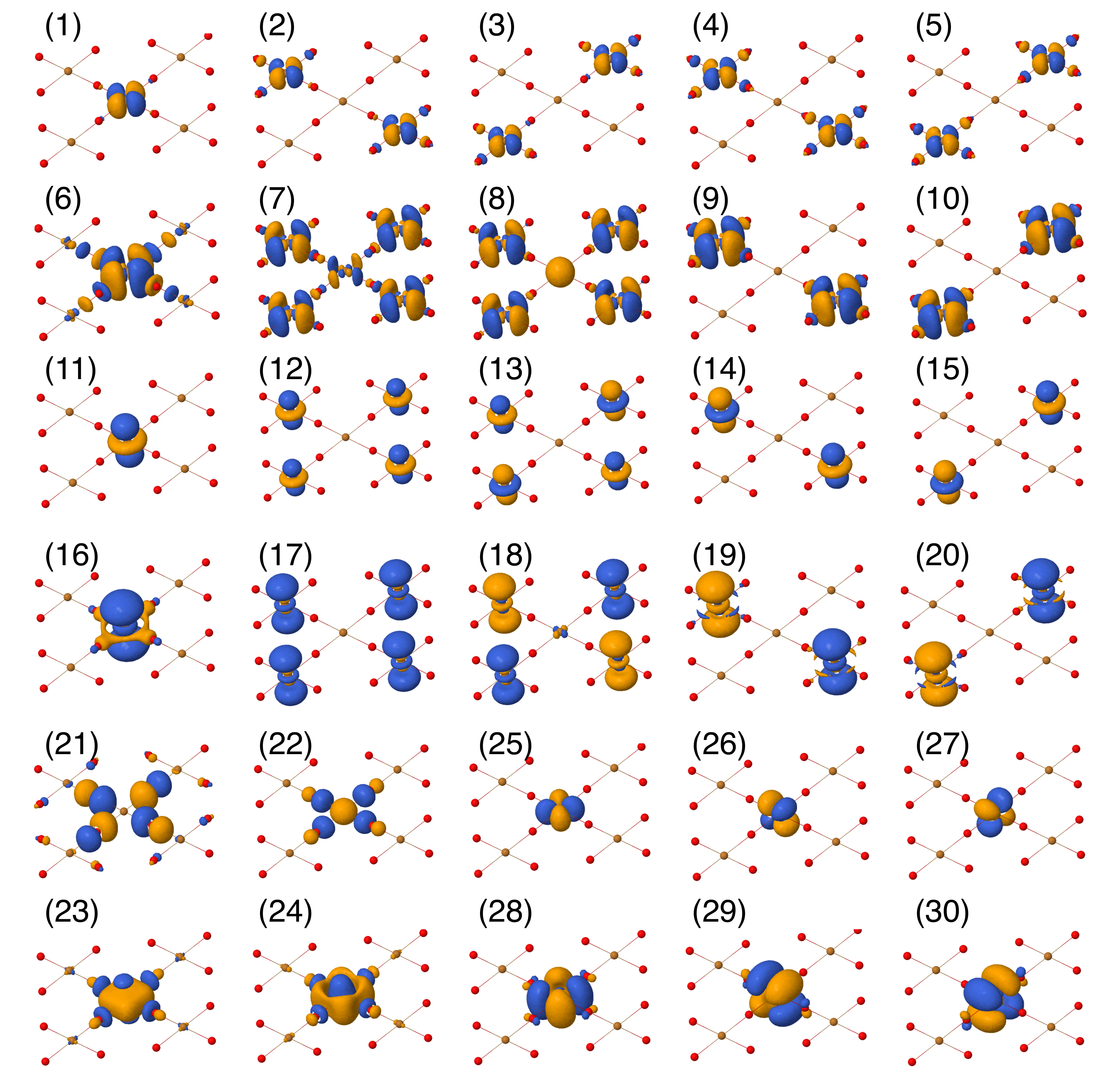}	
		\caption{Active orbital basis used in the CASSCF calculations, 
			plotted using Jmol~\cite{jmol}. 
		}
		\label{activespace_orb}
	\end{center}
\end{figure}

  \begin{table}[!t]
	\caption{The different active spaces (CAS) considered in this work.
		NEL is number of active electrons and NORB is the number of active orbitals.
		The numbers in parenthesis indicate the orbital numbers in Fig~\ref{activespace_orb}. 
	} 
	\label{activespaces}
	\begin{center}
		\begin{tabular}{lcc}
			\hline
			\hline\\
			CAS       & NEL  & NORB  \\
			\hline\\
			CAS-1     & 18   & 24 (1-24)  \\
			CAS-2     & 24   & 30 (25-30) \\
			CAS-3\footnote{The four neighbouring Ni$^{1+}$ (Cu$^{2+}$) ions in the quantum cluster are treated as closed shell Cu$^{1+}$ (Zn$^{2+}$) ions.}
			          & 12   & 14  (1, 6, 11, 16 and 21-30)   \\
			\hline
			\hline
		\end{tabular}
	\end{center}
\end{table}
 To investigate the role of different interactions in the $d^8$ ground state, 
 two different active spaces were considered.
In the first active space, CAS-1 in Table ~\ref{activespaces}, only the orbitals in the $b_1$ and $a_1$ irreps are active. 
These are $d_{x^2-y^2}$ and $d_{z^2}$-like orbitals respectively, and the corresponding double-shell $4d$ orbitals of each of the five Ni (Cu) atoms.
CAS-1 also contains the symmetry-adapted ZR-like composite O 2$p$ and the double-shell 3$p$-like orbitals, numbers 1-20 and 21-24  in Fig.~\ref{activespace_orb}. 
At the mean-field HF level of theory, there are 16 electrons within this set of orbitals, resulting in CAS(16,22) active space.
%, two on the central Ni (Cu), three on each neighbouring Ni (Cu) and two on the ZR-like O 2$p$. 
%We thus write the CASSCF active space as   
In the second active space, CAS-2, orbitals of $b_2$ and the $e$ irreps from the central Ni (Cu) $d$ manifold are also included. 
These are the 3$d_{xy}$, 3$d_{xz,yz}$-like orbitals and the corresponding $4d$ orbitals and the  six electrons, numbers 25-30 in Fig.~\ref{activespace_orb}, resulting in a CAS(24,30) active space. 
The latter active space takes into account the $d^8$ multiplet effects within the $3d$ manifold explicitly.  
%In Fig.~\ref{fig1} the localized symmetry adapted orbitals considered in the active space for the two models are shown. 

The two active spaces considered in this work not only describe all the physical effects included in the above mentioned three-band Hubbard model but go beyond. 
More importantly, we do have any \textit{ad-hoc} input parameters for the calculation as 
all the physical interactions are implicitly included in the {\it ab initio} Hamiltonian describing the actual scenario in the real materials. 
 We employed {\sc OpenMolcas}~\cite{fdez_galvan_openmolcas_2019} quantum chemistry package for all the calculations. 
  
 \section{Results}
\subsection{Ground state of the \boldmath${d^8}$ configuration}

Starting from the electronic structure of the parent compounds, where each Ni (Cu) is in the $d^9$ configuration, we compute the electron-removal (in the photoemission terminology) $d^8$ state to investigate the hole-doped quasiparticle state.  
%For the moment let us assume that the added hole sits on the central Ni (Cu) ion, making it a $d^8$ . 
Since the parent compounds in $d^9$ configuration have strong nearest neighbour antiferromagnetic (AF) correlations~\cite{katukuri_electronic_2020}, the total spin of our QC in undoped case, with five Ni (Cu) sites, in the AF ground state is $S_{QC}=3/2$.  
By introducing an additional hole (or removing an electron) from the central Ni (Cu) in our QC, the $S_{QC}$ values range from 0 to 3.  
To simplify the analysis of the distribution of the additional hole, we keep the spins on the four neighbouring Ni (Cu) sites parallelly aligned in all our calculations and from now on we only specify the spin multiplicity of the central Ni (Cu)O$_4$ plaque. 
The multiplet structure of the $d^8$ configuration thus consists of only spin singlet and triplet states, spanned by the four irreps of the $3d$ manifold. 
The active spaces we consider in this work allow us to compute accurately the excitations only within the $b_1$ and $a_1$ irreps
\footnote{For an accurate quantitative description of the multiplet structure spanned by the other two irreps $b_1$ and $e$, one would need to extend the active space and include the $3d$ and $4d$ manifolds of the four neighbouring Ni (Cu) atoms as well as the O 2$p$ orbitals of the same symmetry, resulting in a gigantic 68 electrons in 74 orbitals active space.}
and we address the full multiplet structure elsewhere.

When computing the  local excitations,  a local singlet state on the central Ni (Cu) corresponds to a total spin on the cluster $S_{QC}=2$. 
However,  a local triplet state, with central spin aligned parallel to the neighboring spins, corresponds to $S_{QC}=3$ and do not satisfy the AF correlations. % in the parent compounds. 
To avoid the spin coupling between the central $d^8$ Ni (Cu) with the neighbouring $d^9$ Ni (Cu) ions, we replace the latter with closed shell, Cu (Zn) $d^{10}$, ions and freeze them at the mean-field HF level. 
Such a simplification is justified, as the local excitation energy we compute is an order of magnitude larger than the exchange interaction~\cite{katukuri_electronic_2020}.   
 %
%The maximum spin $S=3$ corresponds to the case when an electron from a doubly occupied orbital is removed from the the central Ni (Cu) and the two unpaired electrons are parallelly aligned to each other, forming a local triplet favoured by the Hunds' couling, and to the neighbouring spins.  
%The $S=1$ is realised when the local triplet is AF aligned to the neighbours, satisfying the AF exchange as well. 
%$S=2$ occurs when a local singlet, either closed-shell or with anti-parallelly aligned local spins against Hunds' coupling.  

In Table \ref{d8-excit}, the relative energies of the  lowest local spin singlets $^1\!A_{1g}$,  $^1\!B_{1g}$ and spin triplet $^3\!B_{1g}$ states are shown.  
These are obtained from CASSCF + CASPT2 calculations with CAS(12,14) active space (CAS-3 in Table~\ref{activespaces}) which includes the 3$d$ and $4d$ orbitals of the central Ni (Cu) ion and the in-plane O 2$p $ and $3p$ orbitals in the $b_1$ irrep. 
In the CASPT2 calculation, the remaining doubly occupied O $2p$, the central Ni (Cu) $3s$ and $3p$ orbitals  and all the unoccupied virtual orbitals are correlated. 
\begin{table}[!t]
	\caption{Relative energies (in eV) of the electron removal $d^8$ states in \NNO\ and the iso-structural \CCO\ obtained from CAS(12,14)SCF and CASSCF+CASPT2 calculations. 
	} 
	\label{d8-excit}

	\begin{center}
		\begin{tabular}{lccccl}
			\hline
			\hline
			State      & \multicolumn{2}{c}{NdNiO$ _{2} $}  & \multicolumn{2}{c}{CaCuO$ _{2} $}  \\
                       & CASSCF & +CASPT2                   &      CASSCF & +CASPT2 \\
            \hline
			$^1\!A_{1g}$  & 0.00  & 0.00 & 0.00  &  0.00  \\
            $^3\!B_{1g}$  & 1.35  & 1.88 & 2.26  &  2.50  \\
            $^1\!B_{1g}$  & 2.98  & 3.24 & 3.21  &  3.33  \\
			\hline
			\hline
		\end{tabular}
	\end{center}
\end{table}
It can be seen that the ground state is of $^1\!A_{1g}$ symmetry and the lowest triplet excited state, with $^3\!B_{1g}$ symmetry, is around 1.88 eV and 2.5 eV for \NNO\  and \CCO\ respectively. 
The AF magnetic exchange in these two compounds is 76 meV and 208 meV respectively~\cite{katukuri_electronic_2020}, and thus we expect that our simplification of making the neighbouring $d^9$ ions closed shell do not over/underestimate  the excitation energies. 
At the CASSCF level, the $^1\!A_{1g}$-$^3\!B_{1g}$ excitation energy is 1.35 eV in \NNO\ while it is 2.26 eV in \CCO. 
Interestingly, the inclusion of dynamical correlations via the CASPT2 calculation, the $^1\!A_{1g}$ in \NNO\ is stabilized by 0.53 eV compared to $^3\!B_{1g}$ state. 
However, in \CCO, the $^1\!A_{1g}$ state is stabilized by only 0.24 eV. 
This indicates that the dynamical correlations are more active in the $^1\!A_{1g}$ state in \NNO\ than in \CCO. 
We note that the hole excitations within the $3d$ orbitals in the irreps $b_2$ and $e$, calculated with this limited active space (CAS-3) results in energies lower than the $^3\!B_{1g}$ and $^1\!B_{1g}$ states.
However, an accurate description of those states requires an enlarged active space that includes not only the same symmetry oxygen 2$p$ and $3p$ orbitals from the central NiO$_4$ plaque but also the 3$d$, 4$d$ manifold of the neighbouring Ni (Cu) ions, making the active space prohibitively large. 
Here, we concentrate on the analysis of the $^1\!A_{1g}$ ground state and address the complete $d^8$ multiplet spectrum elsewhere. 
\begin{table}[!b]
	\caption{
		Ni and Cu $3d^8$ $^1\!A_{1g}$ ground state wavefunction: Weights (\%) of the leading configurations
		in the wavefunction computed for  \NNO\ and \CCO\ with active spaces CAS-1 and CAS-2 (see Table~\ref{activespaces}).
		$d_{b_1}$ and $p_{b_1}$ are the localized  Ni (Cu) $3d_{x^2-y^2}$ and the oxygen  $2p$ ZR-like orbitals (see Fig.~\ref{fig1}) in the $b_1$ irrep respectively.  
		Arrows in the superscript indicate the spin of the electrons and a $\square$ indicates two holes. %(20 M walkers)
	}
	\begin{center}
		\begin{tabular}{l llll}
			\hline
			\hline\\[-0.30cm]
			  &   \multicolumn{2}{c}{NdNiO$ _{2} $}  & \multicolumn{2}{c}{CaCuO$ _{2} $}  \\ 
			  $^1\!A_{1g}$                         &  CAS-1     & CAS-2   & CAS-1    & CAS-2 \\
			\hline
			\\[-0.20cm]
			$|d_{b_{1}}^\square p_{b_{1}}^{\uparrow \downarrow} \rangle$        &  51.87 & 42.40 &  4.20  & 20.25  \\[0.3cm]
			$|d_{b_{1}}^{\uparrow}p_{b_{1}}^{\downarrow} \rangle$        &   8.27  & 10.48 & 42.58 & 38.52  \\[0.3cm]
			$|d_{b_{1}}^{\downarrow}p_{b_{1}}^{\uparrow} \rangle$        &   6.07  &  7.60  & 25.00 & 25.60  \\[0.3cm]
			$|d_{b_{1}}^{\uparrow \downarrow}p_{b_{1}}^\square \rangle$        &   0.09  &  0.23  & 21.56 &  5.14   \\[0.3cm]

%			$|d_{b_{1}}^2L_{b_{1}}^0d_{b_{1g},}^{nn 1}\rangle$   & x.xx&  xx.x\\
			
			\hline
			\hline
		\end{tabular}
	\end{center}
	\label{wfn}
\end{table}

\subsection{Wavefunction of the electron-removal \boldmath$d^8$ ground state}  
The $^1\!A_{1g}$ ground wavefunction in terms of 
the weights of the four leading configurations (in the case of \CCO) is shown in Table~\ref{wfn}.
The wavefunctions corresponding to the CASSCF calculations with the active spaces CAS-1 and CAS-2 are shown. 
The basis in which the wavefunctions are represented is constructed in two steps:
1) A set of natural orbitals are generated by diagonalising the CASSCF one-body reduced density matrix. 
% However, they obtain the shape resembling the molecular bonding and anti-bonding orbitals.
2) To obtain a set of atomic-like symmetry-adapted localized orbital basis, we localize the Ni (Cu) $3d$ and O $2p$ orbitals on the central NiO$_4$ (CuO$_4$) plaque through a unitary transformation.
Such partial localization within the active space keeps the total energy unchanged. 
The resulting 3\dxtyt\ and the ZR-like oxygen 2$p$ orbital basis is shown in Fig~\ref{fig1}.
FCIQMC calculation was performed in this partial localized basis to obtain the wavefunction as a linear combination of Slater determinants. 
10 million walkers were used to converge the FCIQMC energy to within 0.1 mHartree.   

From Table~\ref{wfn} it can be seen that the electron-removal $d^8$ ground state wavefunction for the two compounds is mostly described by the four configurations spanned by the localized 3\dxtyt\ ($d_{b_1}$) and the symmetry-adapted ZR-like oxygen 2$p$ ($p_{b_1}$) orbitals that are shown in Fig.~\ref{fig1}.
Let us first discus the wavefunction obtain from the CAS-1 active space. 
For \NNO, the dominant configuration involves two holes on 3\dxtyt, $|d_{b_{1}}^\square p_{b_{1}}^{\uparrow \downarrow} \rangle$, and contributes to $\sim$52\% of the wavefunction,
while the configurations that make up the ZR singlet, $|d_{b_{1}}^{\uparrow}p_{b_{1}}^{\downarrow} \rangle$ and $|d_{b_{1}}^{\downarrow}p_{b_{1}}^{\uparrow} \rangle$, contributes to only $\sim$14\%. 
On the other hand, the $d^8$ $^1\!A_{1g}$ state in \CCO\ is predominantly the ZR singlet with $\sim$68\% weight.
In the CASSCF calculation with CAS-2 active space, where all the electrons in the 3$d$ manifold are explicitly correlated,  
we find that the character of the wavefunction remains unchanged in \NNO\ but weight on the dominant configurations is slightly reduced. 
On the other hand, in \CCO, while the contribution from the ZR singlet is slightly reduced, the contribution from $|d_{b_{1}}^\square p_{b_{1}}^{\uparrow \downarrow} \rangle$ configuration is dramatically increased at the expense of the weight on 
$|d_{b_{1}}^{\uparrow \downarrow}p_{b_{1}}^\square \rangle$. 
This demonstrates that the additional freedom provided by the $d_{xy}$ and $d_{xz/yz}$ orbitals for the electron correlation helps to accommodate the additional hole on the Cu ion.

We note that the four configurations shown in Table~\ref{wfn} encompass almost 90\% of the $d^8$ wavefunction (with CAS-2 active space) in \CCO. 
Thus, the use of a three-band Hubbard model~\cite{emery_3b_hubbard_prl_1987,jiang_cuprates_prb_2020} to investigate the role of doped holes in CuO$_2$ planes is a reasonable choice. 
However, for \NNO\ these configurations cover only 60\% of the $d^8$ wavefunction, hence a three-band Hubbard model is too simple to describe the hole-doped monovalent nickelates. 

\begin{figure}[!t]
\begin{center}
	\includegraphics[width=0.48\textwidth]{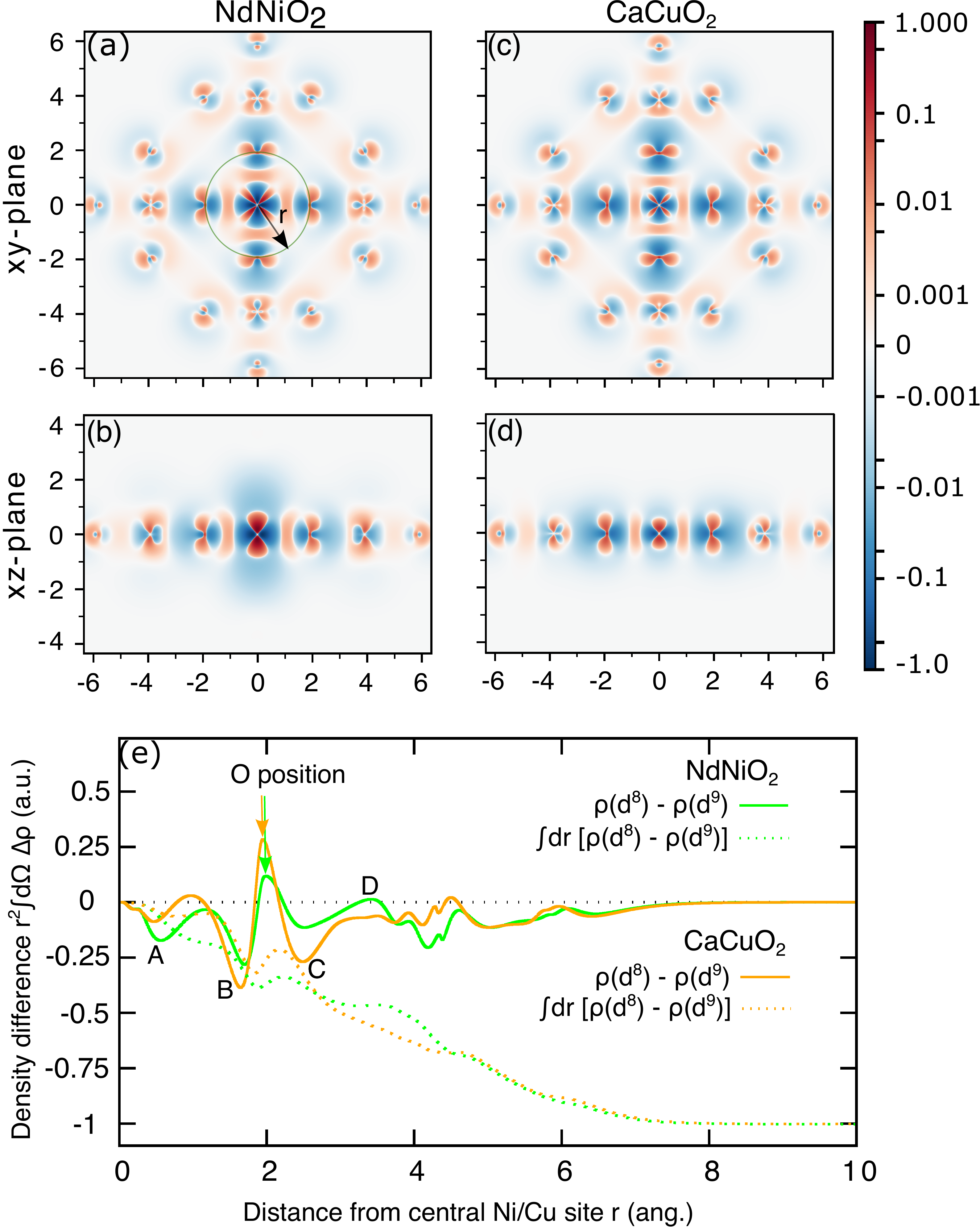}	
	\caption{Electron density difference of the $d^8$ and $d^9$ ground states ($\rho(d^8) - \rho(d^9)$) for \NNO\ in the xy-plane (a) and xz-plane (b),  and for \CCO\ xy-plane (c) and xz-plane (d). 
	The coordinates of the central Ni (Cu) $d^8$ ion are set to (0,0). The scale of the heat-bar is logarithmic between $\pm$0.001 to $\pm$1.0 and is linear between 0 and $\pm$0.001. 
	(e) Electron density difference integrated over a sphere centered on on the central Ni(Cu) atoms (full curves) as a function of the radius $r$ shown in (a). 
	The result of an additional radial integration (dashed curves) as a function of the upper integration limit.}
	\label{fig2}
\end{center}
\end{figure}

A more intuitive and visual understanding of the distribution of the additional hole can be obtained by plotting the difference of the $d^8$ and the $d^9$ ground state electron densities as shown in Fig.~\ref{fig2}. 
Electron density of a multi-configurational state can be computed as a sum of densities arising from the natural orbitals and corresponding (well-defined) occupation numbers.
We used Multiwfn program \cite{Multiwfn} to perform this summation.
The negative values of the heat map of the electron density difference (blue color) and the positive values (in red) represent respectively the extra hole density and additional electron density in $d^8$ state compared to the $d^9$ state.
From Fig.~\ref{fig2}(a)/(c) that show the density difference in the NiO$_2$/CuO$_2$ planes (xy-plane), we conclude the following: 
\begin{enumerate}
\item The hole density is concentrated on the Ni site (darker blue) with $b_1$ (\dxtyt) symmetry in \NNO\ whereas 
 it is distributed evenly on the four oxygen and the central Cu ions with $b_1$ symmetry in \CCO, a result consistent with the wavefunction reported in Table~\ref{wfn}.
\item In \NNO, the hole density is spread out around the Ni ion with larger radius, and otherwise in \CCO.  
 This demonstrates that the $3d$ manifold in Cu is much more localized than in Ni and therefore the onsite Coulomb repulsion $U$ is comparatively smaller for Ni.
\item  The darker red regions around the Ni site in \NNO\ indicate stronger $d^8$ multiplet effects that result in rearrangement of electron density compared to $d^9$ configuration.
\item  In \CCO, we see darker red regions on the oxygen ions instead, which shows that the significant presence of a hole on these ions results in noticeable electron redistribution. 
\end{enumerate}

The electron density difference in the xz-plane (which is perpendicular to the NiO$_2$/CuO$_2$ planes) is quite different in the two compounds. 
The hole density in \NNO\ is spread out up to 2\,\AA\ in the $z$-direction, unlike in \CCO, where it is confined to within 1\,\AA .
We attribute this to the strong radial-type correlations in \NNO. 
With the creation of additional hole on the 3\dxtyt\ orbital, the electron density which is spread out in the \dzt\ symmetry via the dynamical correlation between 3\dzt\ and 4\dzt\ orbitals~\cite{katukuri_electronic_2020}, becomes more compact in the \dzt\ symmetry through the reverse breathing.  
Thus, we see a strong red region with 3\dzt\ profile and a blue region with expanded 4\dzt\ profile. 

To obtain a quantitative understanding of the charge density differences for the two compounds, in Fig.~\ref{fig2}(e) we plot the electron density difference integrated over a sphere centered on the central Ni(Cu) atom as a function of the radius $r$ shown in Fig.~\ref{fig2}(a).
Four features, which we marked A-D, clearly demonstrate the contrast in the charge density differences in the two compounds. 
From the feature A at $r$ close to Ni (Cu), it is evident that
% the hole density on Ni in \NNO\ is larger than on Cu in \CCO\ and
% the extent of hole density around Ni (Cu) is larger (compact) in \NNO\ (\CCO).
the extent of hole density around Ni in \NNO\ is larger than around Cu in \CCO.
The features B and C that are on either side of the position of oxygen ions show that the hole density is significantly larger on oxygen atoms in \CCO\ than in the \NNO.
It is interesting to note that we see a jump (feature D) in the electron density above zero at $r$ close to the position of Nd ions in \NNO, while in \CCO\ the curve is flat in the region of Ca ions. 
This shows that there is some electron redistribution happening around the Nd ions. 

The hole density within a solid sphere (SS) around the central Ni (Cu) atom obtained by additional integration over the radius $r$ is also shown in Fig.~\ref{fig2}(e) with dashed curves. 
It can be seen that the total hole density within the SS of $r\sim$4\,\AA, where the neighboring Ni (Cu) ions are located, is only $\sim$0.5 in both the compounds, with slight differences related to the feature D. 
This is due to the screening of the hole with the electron density pulled in from the farther surroundings. 
% On a closer look, the hole density in SS with $r\sim$4\,\AA in \CCO\ is larger and 0.5, indicating that the screening is relatively smaller in \CCO\ than in \NNO. 
As one would expect, a SS with $r$ of the size of the cluster, the total hole density is one in both the compounds. 

\begin{figure}[!b]
	\begin{center}
		\includegraphics[width=0.480\textwidth]{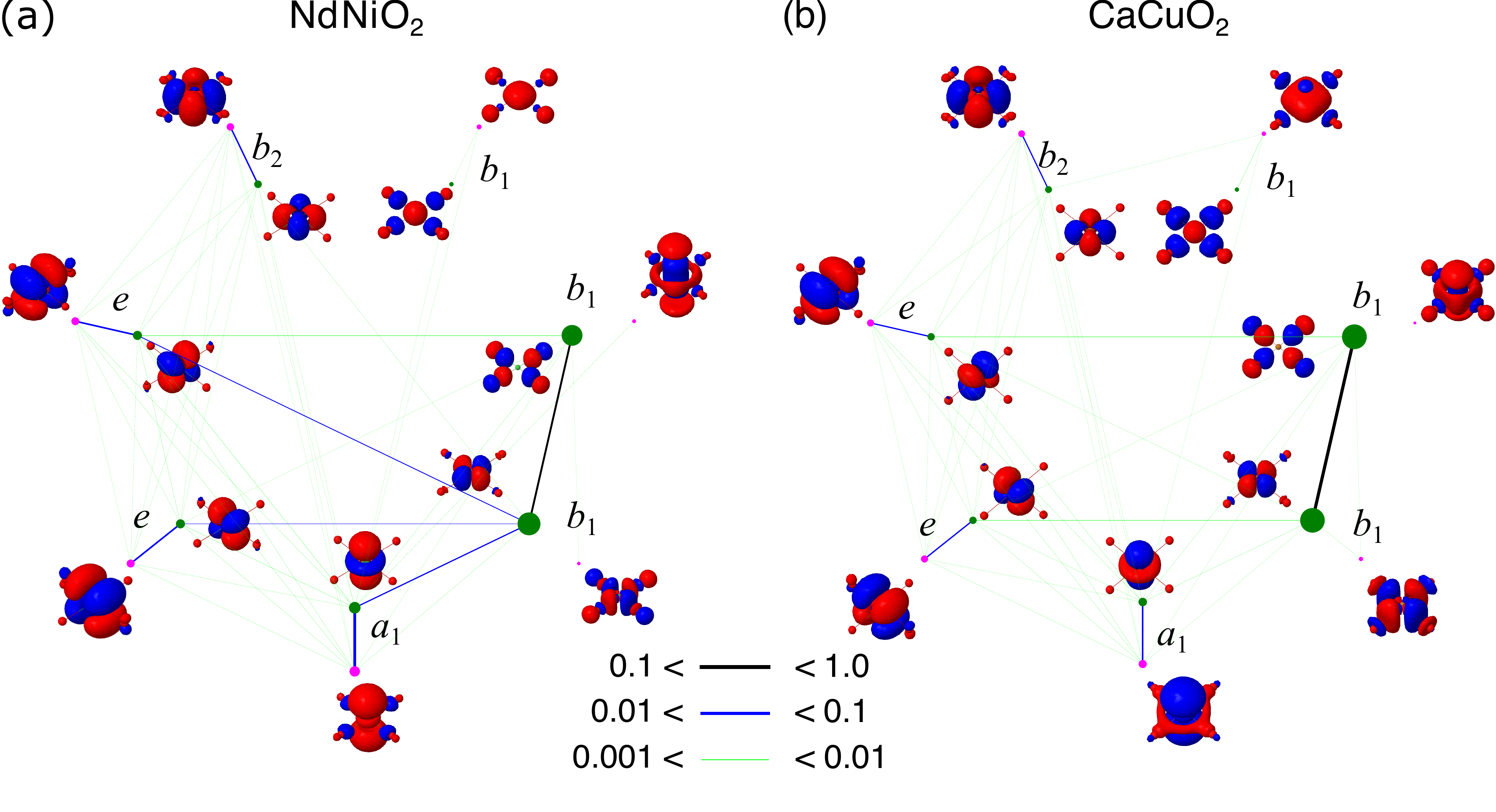}	
		\caption{Single orbital entanglement entropy, $s(1)_i$, (dots) and mutual orbital entanglement entropy, $I_{i,j}$, (colored lines) of the orbital basis used to expand the $d^8$ wavefunction in Table~\ref{wfn} for \NNO\ (a) and \CCO\ (b). 
		Entanglement entropy of the orbitals centred on the central NiO$_4$/CuO$_4$ plaque are only shown. 
		The irrep to which the orbitals belong to are also shown. 
		The green and magenta colors represent the two different set of orbitals, occupied (at the HF level) and the corresponding double-shell (virtual), respectively. 
		The thickness of the black, blue and green lines denote the strength of $I_{i,j}$, and the size of the dots is proportional to $s(1)_i$.
		}
		\label{entanglement}
	\end{center}
\end{figure}

\subsection{Orbital entanglement entropy }
To analyse the different type of correlations active in the two compounds in $d^8$ configuration, we compute the entanglement entropy~\cite{boguslawski_entanglement_2012,boguslawski_orbital_2013,boguslawski_orbital_2015}.  
While the single orbital entropy, $s(1)_i $, quantifies the correlation between $i$-th orbital and the remaining set of orbitals, 
the mutual information,  $I_{i,j}$ is the two-orbital entropy between $i$ and $j$~\cite{legeza_optimizing_2003,rissler_measuring_2006}, and illustrates the correlation of an orbital with another, in the embedded environment comprising of all other orbitals. 
We used {\sc QCMaquis}~\cite{keller_an_2015} embedded in {\sc OpenMolcas}~\cite{fdez_galvan_openmolcas_2019} package to compute the entropies.

In Figure~\ref{entanglement}, $s(1)_i$ and $I_{i,j}$ extracted from CASSCF calculations with CAS-2 active space for \NNO\ and \CCO\ are shown. 
The orbital basis for which the entropy is computed is the same as the basis in which the wavefunction presented in Table~\ref{wfn} is expanded.  
As mentioned previously, this orbital basis is obtained from partial localization of the natural orbitals in a way that only the 3\dxtyt\ and the O 2$p$ ZR-like orbitals are localized. 
Since a large part of electron correlation is compressed in natural orbitals, we see a tiny $s(1)_i$ for all orbitals except for the localized  3\dxtyt\ and the O 2$p$ ZR-like orbitals where it is significant. This is consistent with the wavefunction in Table~\ref{wfn}.
The mutual orbital entanglement between pairs of orbitals shows strong entanglement between the  3\dxtyt\ and the O 2$p$ ZR-like orbitals for both \NNO\ and \CCO, a consequence of the dominant weight of the configurations spanned by these two orbitals in the wavefunction. 
The next strongest entanglement is between the Ni/Cu 3$d$ valence and their double-shell $4d$ orbitals.
Such strong entanglement also observed for the undoped $d^9$ ground state~\cite{katukuri_electronic_2020}, is a result of dynamical radial correlation \cite{helgaker_molecular_2000} and orbital breathing effects~\cite{gunnarsson_density-functional_1989,bogdanov_natphys_2021}. 
Interestingly, the entanglement entropy in the range 0.001-0.01 (green lines) is quite similar in the two compounds, although one sees more entanglement connections in \NNO.  
A comparison of the entropy information between \NNO\ and \CCO\ reveals that the Ni 3$d$ and 4$d$-like orbitals contribute rather significantly (thicker blue lines) to the total entropy, in contrast to the Cu 3$d$ and 4$d$-like orbitals, something that is also seen in the undoped compounds~\cite{katukuri_electronic_2020}. 

\section{Conclusions and discussion}
In conclusion, 
our {\it ab initio} many-body quantum chemistry calculations for the electron removal ($d^8$) states find a low-spin closed-shell singlet ground state in \NNO\ and  that the additional hole is mainly localized on the Ni 3\dxtyt\ orbital, unlike in CaCuO$_2$, where a Zhang-Rice singlet is predominant. 
We emphasise that the $d^8$ wavefunction is highly multi-configurational where the dominant closed-shell singlet configuration weight is only $\sim$42\%.
This result is consistent with the experimental evidence~\cite{rossi2020orbital,goodge-a} of orbitally polarized singlet state as well as the presence of holes on the O $2p$ orbitals.
Importantly, the persistent dynamic radial-type correlations within the Ni $d$ manifold result in stronger $d^8$ multiplet effects in \NNO, and consequently the additional hole foot-print is more three dimensional. 
In \CCO, we find that the electron correlations within the $d_{xy}$ and $d_{xz/yz}$ orbitals changes the hole-doped wavefunction significantly. Specifically, the double hole occupation of Cu \dxtyt\ is significantly increased and this can influence the transport properties. 

It was recently proposed that nickelates could be a legitimate realization of the single-band Hubbard model~\cite{kitatani_nickelate_2020}. 
However, our analysis shows that even the three-band Hubbard model~\cite{eskes1991a}, which successfully describes the hole-doped scenario in cuprates, falls short to describe hole-doped nickelates and additional orbital degrees of freedom are indeed necessary for the description of the strong multiplet effects we find.
Much has been discussed about the importance of  rare-earth atoms for the electronic structure of superconducting nickelates, e.g. see~\cite{nomura2021superconductivity}. 
The three-dimensional nature of the hole density we find in \NNO\ might also be hinting at the importance of out-of-plane Nd ions.  
It would be interesting to compare the hole density of \NNO\ with other iso-structural nickelates such as \LNO\ where La $5d$ states are far from the Fermi energy.  
Since the infinite-layered monovalent nickelates are thin films and often grown on substrates, one could ask the question of how the electronic structure of the undoped and doped compounds changes with varying Ni-O bond length. Would this influence the role of electronic correlations in $d^9$ nickelates? We will address these in the near future. 

\section*{Conflict of Interest Statement}
The authors express no conflict of interests. 

\section*{Author Contributions}
VMK and AA designed the project. VMK and NAB performed the calculations. All the authors analysed the data. VMK wrote the paper with inputs from NAB and AA. 

\section*{Funding}
We gratefully acknowledge the Max Plank Society for financial support. 

\section*{Acknowledgments}
VMK would like to acknowledge Giovanni Li Manni and Oskar Weser for fruitful discussions.

%\bibliographystyle{unsrt}
%\bibliography{ref,nno_ref,nno_ref_2,sc_nickelates,cuprates,codes,cobaltates,scwto,mypapers,basis_sets,fciqmc}

%%% Make sure to upload the bib file along with the tex file and PDF
%%% Please see the test.bib file for some examples of references

\end{document}